\documentclass[12pt]{article}
\title{Zonal density staircase formation in collisional drift-wave turbulence}
\author{M. Leconte$^1$ and T. Kobayashi$^{2,3}$ \\
{$^1$ Korea Institute of Fusion Energy (KFE), Daejeon 34133, South Korea} \\
$^2$ National Institute for Fusion Science, National Institutes \\ of Natural Sciences, 
Toki 509-5292, Japan \\
$^3$ The Graduate University for Advanced Studies, SOKENDAI, \\ Toki, 509-5292, Japan \\\quad\\
Email: mleconte@kfe.re.kr}

\usepackage{amsmath}
\usepackage{graphicx}

\newcommand{\dif}{\partial}
\newcommand{\wk}{\omega_k}
\newcommand{\kx}{k_x}
\newcommand{\ky}{k_y}

\newcommand{\gl}{\gamma_L}
\newcommand{\dia}{\omega_{*0}}
\newcommand{\cp}{\theta_k}

\newcommand{\anorm}{\hat \alpha}

\newcommand{\zv}{U}
\newcommand{\zn}{N}

\begin{document}
\maketitle

\begin{abstract}
Turbulence-driven quasi-stationnary structures known as 'staircase' are investigated using the collisional drift-wave model. Two-dimensional simulations show that the ability of zonal density corrugations to suppress turbulence are affected by the adiabaticity parameter (inversely proportional to collision frequency). As the adiabaticity parameter increases, zonal density becomes less efficient at suppressing turbulence, and zonal flows become dominant in the near-adiabatic regime. The nonlinear transport crossphase displays radial modulations associated to zonal density.
\end{abstract}


\section{Introduction}
The High-confinement regime (H-mode) is important for future fusion devices like ITER. It has been the focus of research for more than 30 years. See e.g. Refs \cite{ConnorWilson2000,Burrell2020, Kobayashi2020} for a review. The presence of turbulence-driven flows, i.e. zonal flows (ZF) have been shown to facilitate access to H-mode, by shearing apart turbulence eddies \cite{DiamondIIH2005}.
Contrary to models based on waves in random media, zonal flows are not random spatially, but instead form well-defined patterns, similar to those found in other non-equilibrium systems \cite{CrossGreenside2009}. The most common radial pattern first observed in gyrokinetic simulations of ion-temperature gradient driven (ITG) turbulence has been dubbed `$E \times B$ staircase' \cite{DifPradalierHornungGhendrih2015, DifPradalierHornungGarbet2017, KosugaDiamond2014,YanDiamond2022}, due to its quasi-periodic nature, for which the leading explanation is that zonal flows are responsible for the pattern, and that the density and temperature profile corrugations and hence transport modulation are a consequence of the zonal flow pattern directly suppressing the turbulence intensity \cite{AshourvanDiamond2017,GuoDiamondHughes2019}.
Similar patterns were observed in the KSTAR tokamak and reproduced by global $\delta f$ gyrokinetic simulations of collisionless trapped-electron modes (CTEM) \cite{Choi2019,LeiKwonHahm2019,QiChoiKwonHahm2021}.
However, it is well-known that turbulent transport does not only depend on the turbulence intensity but also on the phase-angle, i.e. transport crossphase \cite{TynanFujisawaMcKee2009, Kobayashi2017, Camargo1995} between electric potential and the advected quantity driving the turbulence, e.g. density, ion temperature, electron temperature, etc \ldots
For particle transport - on which we focus here - the turbulent particle flux can be written in the form:
$\Gamma = \displaystyle \sum_k k_y \sqrt{|n_k|^2} \sqrt{|\phi_k|^2} \gamma_{\rm coh}^2 \sin \cp$ \cite{Kobayashi2017}. Here, $|n_k|^2$ and $|\phi_k|^2$ denote the power spectrum of density and potential fluctuations, respectively, $\gamma_{\rm coh}$ is the coherence, assumed here to be $\gamma_{\rm coh} \simeq 1$ for simplicity, and $\cp$ is the transport crossphase (crossphase spectrum), i.e. the phase-angle between density and potential fluctuations.
In most research works, it is often assumed that the transport crossphase between density and potential is linear, leading to the so-called `$i \delta$' prescription. There are some exceptions, e.g. Refs \cite{Terry2003, WareTerryDiamond1996}. In gyrokinetic simulations, the nonlinear crossphase in wavenumber space appears to closely match its linear value, supporting the $i \delta$ prescription for ITG and TEM turbulence \cite{ToldJenkoGorler2013} . Gyrokinetic simulations of collisionless trapped-electron mode \cite{LangParkerChen2008} revealed that - in certain parameter regimes e.g. cold ions relative to electrons - zonal flows are ineffective at suppressing turbulence, and instead zonal density generation becomes the dominant saturation mechanism.
In previous work, one of the author (M.L), showed that zonal flows can affect the transport crossphase, in the framework of a parametric instability analysis of a fluid model for dissipative trapped-electron mode \cite{LeconteSingh2019}. Additionally, an equation was derived for the dynamics of zonal density amplitude. Ref. \cite{RameswarSinghDiamond2021} derived an equation for zonal density staircase evolution and for the correlation between zonal density and zonal flows. In \cite{LeconteKobayashi2021}, we presented numerical results showing that the transport crossphase might in fact be nonlinear. This nonlinearity manifests itself mostly via radial modulations of the crossphase, not predicted by linear theory. This radial modulation is responsible for the generation of zonal density corrugations. Such corrugations of the density profile were observed in Ref. \cite{Kobayashi2014} during limit cycle oscillations preceding the L-H transition in JFT-2M, using the heavy ion beam probe diagnostic (HIBP). We present here an extended study of zonal density staircase formation and crossphase modulations. The effect of the adiabaticity parameter - inversely proportional to collisionality - on the ability of zonal density corrugations to suppress turbulence is investigated.
The article is organized as follows: In section 2, we present the extended wave-kinetic model for collisional drift-waves, including density profile corrugations (zonal density). In section 3, we identify the associated energy transfers. In section 4, we present numerical results of 2D drift-fluid simulations and compare with the analytical model of section 2.
The results are discussed in section 5 and finally we present a conclusion.

\section{Model}
We analyse the 2 field modified Hasegawa-Wakatani model, a basic representative model for edge turbulence in magnetized plasmas \cite{WakataniHasegawa1984, Numata2007,StoltzfusDueck2016}:
\begin{eqnarray}
\frac{\dif n}{\dif t} + \{ \phi, n\} + \kappa \frac{\dif \phi}{\dif y} & = & - \alpha ( \tilde n - \tilde \phi),
\label{hw1} \\
\frac{\dif \nabla_\perp^2 \phi }{\dif t}  + \{ \phi, \nabla_\perp^2 \phi \} & = & - \alpha (\tilde n - \tilde \phi),
\label{hw2}
\end{eqnarray}
where $n$ is the electron density and $\phi$ is the electric potential, $\{ f , g \} = \frac{\dif f}{\dif x} \frac{\dif g}{\dif y}  - \frac{\dif f}{\dif y} \frac{\dif g}{\dif x}$ denote Poisson brackets. Here, $\tilde n = n - \langle n \rangle$ and $\tilde \phi = \phi - \langle \phi \rangle$ denote the non-zonal components of the fields, and $\langle \ldots \rangle = (1/L_y) \int \ldots dy$ is the zonal average. The quantity $\kappa$ is the normalized density-gradient, and  $\alpha = k_\parallel^2 v_{Te}^2 / \nu_{ei}$ is the coupling parameter, with $k_\parallel \sim 1 / (qR)$ the parallel wavenumber, $v_{Te} = \sqrt{T_e / m_e}$ the electron thermal velocity and $\nu_{ei}$ the electron-ion collision frequency. Note that an important control parameter of the Hasegawa-Wakatani model  is the \emph{adiabaticity parameter}: the ratio $\anorm = \alpha/ \kappa$.
Time and space are normalized as: $\omega_{c,i} t \to t$ and $\rho_s \nabla_\perp \to \nabla_\perp$, with $\rho_s = c_s / \omega_{c,i}$ the sound gyroradius, $c_s = \sqrt{T_e / m_i}$ the sound speed, and $\omega_{c,i} = eB / m_i$ the ion Larmor frequency.

We extend the wave-kinetic model of Sakaki \emph{et al.} \cite{Sasaki2018}, to include zonal profile corrugations, i.e. zonal density.
After some algebra, one obtains the following reduced model:
\begin{eqnarray}
\frac{\dif W_k}{\dif t} + \frac{\dif \wk}{\dif \kx} \nabla_x W_k - \ky \nabla_x \zv \frac{\dif W_k}{\dif \kx} & = & 2 \gl W_k -  2 c_k W_k \nabla_x \zn
- \Delta\omega W_k^2,
\quad \label{wke1} \\
\frac{\dif \zv}{\dif t} & = & - \nabla_x \Pi + \nu_\perp \nabla_{xx} \zv - \mu \zv,
\label{zv1} \\
\frac{\dif \zn}{\dif t} & = & - \nabla_x \Gamma + \nabla_x \Big[ (D_0+D_t(W_k)) \nabla_x \zn \Big],
\qquad \label{zn1}
\end{eqnarray}
Details of the derivation are given in Appendix. \\
Eq. (\ref{wke1}) is the extended wave kinetic equation (EWKE) for the wave action density $W_k$ , which includes a nonlinear contribution to the growth-rate, due to the zonal density induced modulation of the transport crossphase, second term on the r.h.s. of Eq. (\ref{wke1}). Eqs. (\ref{zv1}) and (\ref{zn1}) describe the dynamics of zonal flows $\zv$ and zonal density corrugations $\zn$, respectively.

\begin{figure}
\begin{center}
\includegraphics[width=0.5\linewidth]{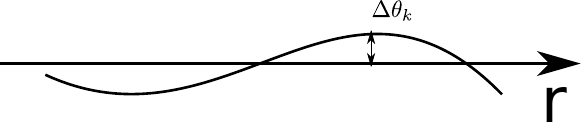}
\end{center}
\caption{Radial modulation $\Delta \cp$ of the transport crossphase, due to zonal profile corrugations.}
\label{radmod}
\end{figure}

\begin{figure}
\begin{center}
\includegraphics[width=0.5\linewidth]{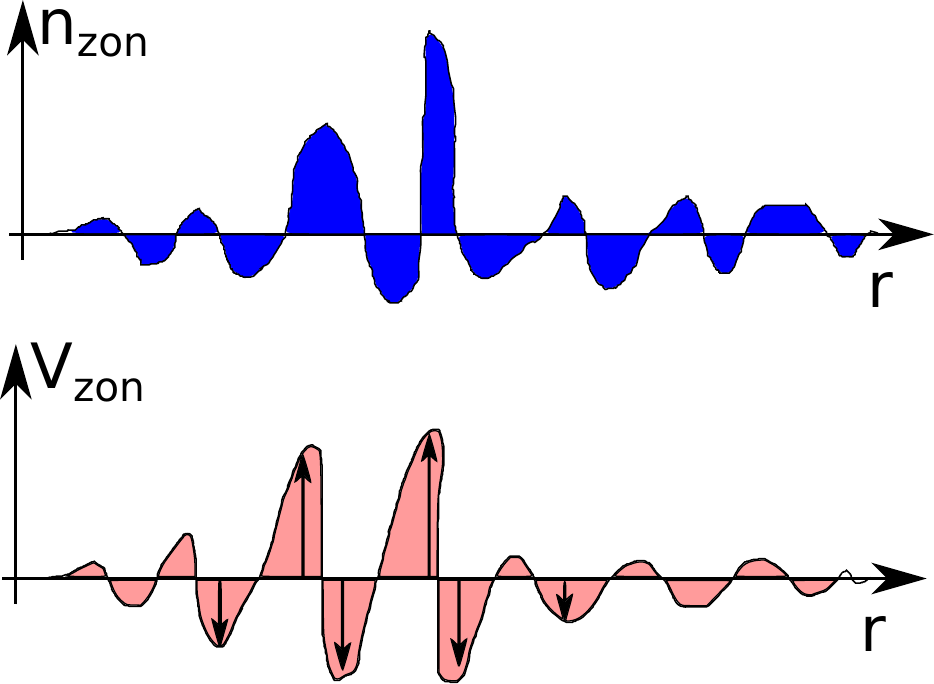}
\end{center}
\caption{Radial modes: sketch of zonal density (top) and zonal flows (bottom).}
\label{zonprof}
\end{figure}

The frequency $\wk$ is the nonlinear frequency, including Doppler-shift due to zonal flows:
\begin{equation}
\wk = \frac{\dia}{1+ k_\perp^2} + \ky \zv, 
\end{equation}
where $k_\perp^2 = \kx^2 + \ky^2$,
and $W_k$ denotes the wave-action density for drift-waves with adiabatic electrons, given by:
\begin{equation}
W_k = (1+ k_\perp^2)^2 |\phi_k|^2,
\end{equation}
where $|\phi_k|^2$ the turbulent power spectrum, and $D_t(W_k) \propto W_k$ is the turbulent diffusivity, while $D_0$ includes residual turbulence and neoclassical contributions.
Moreover, the nonlinear growth-rate, due to non-adiabatic electrons is:
\begin{equation}
\gamma_{k} = \gl -c_k \nabla_x \zn,
\end{equation}
with $\gl = \cp^0 \wk^L$ the linear growth-rate, $\cp^0$ the linear crossphase between density $n_k$ and potential $\phi_k$, e.g. $\cp^0 \sim (\dia - \wk^L) / \alpha$ for collisional drift-waves $\alpha \gg 1$ and $c_k = c_s k_y \gl / \dia = c_s k_y \cp^0 / (1+k_\perp^2)$, with $c_s$ the sound-speed. Note that for linear drift-waves, $n_k^L = (1- i \cp^0) \phi_k$, but nonlinearly $n_k \simeq (1-i \cp^0 -i \Delta \cp(x,t))$, with $\Delta \cp / \cp^0 = - \nabla_x \zn / |\nabla_x n_0|$, where $\nabla_x n_0 <0$ is the equilibrium density gradient. Physically, this means that the transport crossphase is \emph{nonlinear} - due to the $E \times B$ convective nonlinearity - and can be viewed as a radial modulation of the crossphase due to zonal profile corrugations, e.g. zonal density. This physical mechanism is sketched [Fig \ref{radmod}], together with a sketch of zonal flows and zonal density [Fig \ref{zonprof}]. The extended wave-kinetic model (\ref{wke1},\ref{zv1},\ref{zn1}) was implemented numerically in Ref. \cite{Sasaki2021}. However, the authors assumed a linear crossphase. Hence, radial modulations of the crossphase are neglected in Ref. \cite{Sasaki2021}.

We now describe the model. Here, the first term  on the r.h.s. of Eq. (\ref{wke1}) is the turbulence drive with linear growth-rate $\gl({\bf k})$, the second term on the r.h.s. is the nonlinear contribution to the growth-rate, proportional to the zonal density gradient $\nabla_x \zn$, where $c_k$ is the $k$-dependent proportionality coefficient. The first term on the r.h.s. of Eq. (\ref{zv1}) is the Reynolds torque which involves the Reynolds stress $\Pi = \langle v_x v_y \rangle$. The first term on the r.h.s. of Eq. (\ref{zn1}) is the convective particle flux associated to the linear electron response $\Gamma = \langle \tilde v_x \tilde n^L \rangle = \sum_k (i c_s \ky/2) [n_k^L \phi_k^* -(n_k^L)^* \phi_k]$.

The Reynolds stress can be expressed in the form:
\begin{equation}
\Pi = - \sum_{\ky} \int \frac{\ky \kx W_k }{(1+ k_\perp^2)^2} d \kx
\end{equation}
Moreover, the particle flux can be approximated as:
\begin{equation}
\Gamma \simeq c_s \sum_{\ky} \int \frac{  \ky \cp^0 W_k }{(1+k_\perp^2)^2} d \kx
\end{equation}
with $\cp^0 = \gl / \wk^L$ the linear crossphase, with $\wk^L = \omega_*^0 / (1+ k_\perp^2)$ the linear DW frequency, and the approximation $\sin \cp^0 \simeq \cp^0$ was used, since we assume $|\cp^0| \ll 1$.


\section{Energy transfer}
Multiplying Eq. (\ref{wke1}) by $(1+ k_\perp^2)^{-1}$ and integrating in $\kx$, one obtains the evolution equation for turbulence intensity $I = \int (1+ k_\perp^2) |\phi_k|^2 d \kx$. Multiplying Eq. (\ref{zv1}) by $2 \zv$ yields the evolution equation for zonal flow intensity $\zv^2$, and multiplying Eq. (\ref{zn1}) by $2 \zn$ yields the evolution equation for zonal density intensity $\zn^2$. This yields the following system:
\begin{eqnarray}
\frac{\dif I}{\dif t} + \frac{\dif}{\dif x} (\hat v_g I) & = & 2 \hat \gl I +W_{turb}^V + W_{turb}^n - \Delta \hat \omega I^2,
\label{wke2} \\
\frac{\dif \zv^2}{\dif t} & = & W_V  - \mu \zv^2 + \nu_\perp \zv \nabla_{xx} \zv,
\label{zv2} \\
\frac{\dif \zn^2}{\dif t} & = & W_n + \zn \nabla_x \left[ (D_0+D_t(I)) \nabla_x \zn \right],
\label{zn2}
\end{eqnarray}
with the nonlinear transfer terms given by:
\begin{eqnarray}
W_{turb}^V & = & -2 \Pi \nabla_x \zv, \\
W_{turb}^n & = & - 2 \hat c I \nabla_x \zn  = - 2 \Gamma \nabla_x \zn,  \\
W_V & = & -2 \zv \nabla_x \Pi, \\
W_n & = & - 2 \zn \nabla_x \Gamma,
\end{eqnarray}
where $\hat c$ is defined via:
\begin{eqnarray}
\hat c I & = & \sum_{\ky} \int c_s k_y \cp^0 W_k (1+ k_\perp^2)^{-2} d \kx, \nonumber\\
 & = & \Gamma,
\end{eqnarray}
Note that the energy is conserved in the turbulence - zonal density interaction, since $\int (W_{turb}^n + W_n) dx = 0$. We stress out that this arises independently from the well-known energy conservation in the turbulence - zonal flow interaction $\int (W_{turb}^V +W_V) dx=0$. This is remarkable as it opens the way to \emph{transport decoupling} in more sophisticated models, e.g. the possibility of different magnitude of profile corrugations for different channels such as particle transport channel and thermal transport channel.


\section{Evidence of the proposed mechanism in numerical simulations of HW turbulence}

\subsection{Numerical results} 

\begin{figure}
\begin{center}
\includegraphics[width=0.5\linewidth]{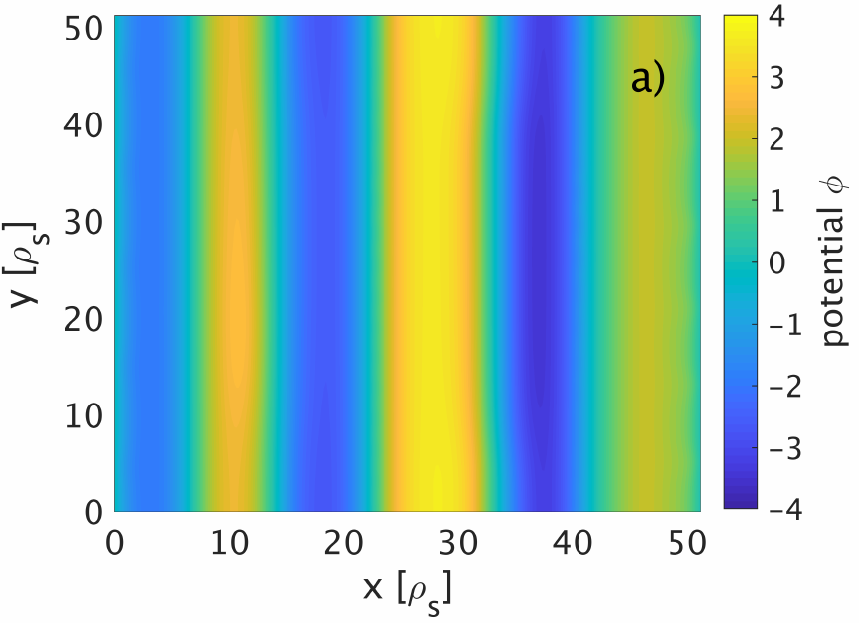}\includegraphics[width=0.5\linewidth]{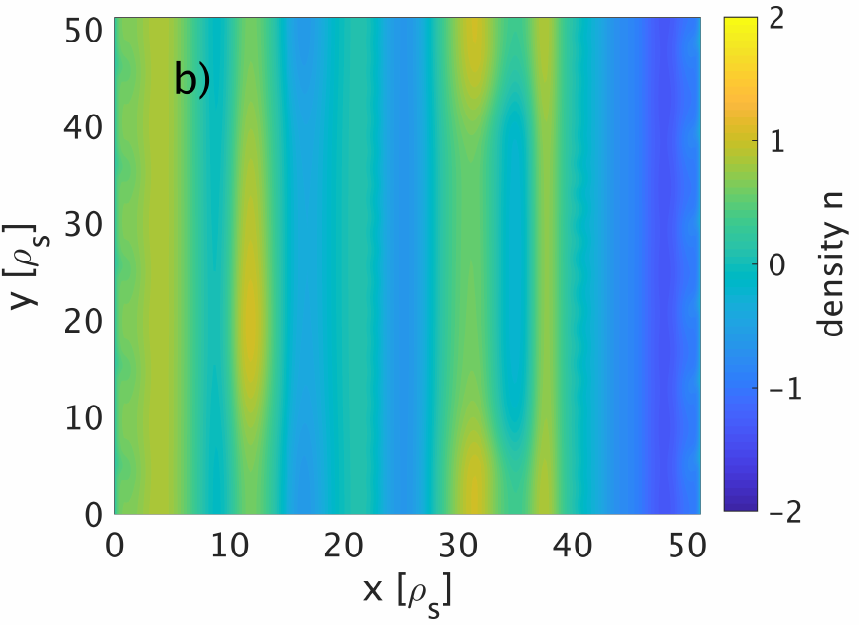}
\caption{Snapshots of a) potential and b) density in the saturated state ($t=10000$), including zonal components.}
\label{fig-snapshot}
\end{center}
\end{figure}

Since we want to test whether radial modulations of the transport crossphase are generated by turbulence, we perform fluid simulations of collisional drift-wave turbulence described by the Hasegawa-Wakatani model  (\ref{hw1}, \ref{hw2}) using the BOUT++ framework \cite{Dudson2009}, employing PVODE with adaptative time stepping to advance in time.
The model used \cite{footn} is 2D with a resolution of $256^2$, integrated over a square of length $L=51.2$. The coupling parameter is set to $\alpha=1$ (unless stated otherwise), and the equilibrium density gradient is $\kappa=0.5$. For Poisson brackets, the Arakawa scheme is used \cite{Arakawa1966}. For numerical stability reasons, viscous hyperdiffusion and particle hyperdiffusion terms are added to the r.h.s. of Eqs. (\ref{hw1}) and (\ref{hw2}) , respectively, with coefficients: $D_\Omega=D_n=1 \times 10^{-4}$.
Simulations are carried out until the turbulence saturates and a statistically stationary state is reached. Snapshots of potential contour [Fig.\ref{fig-snapshot}a] and density contour [Fig.\ref{fig-snapshot}b] are shown, including zonal components. Contours of potential and density are elongated in the poloidal direction $y$ due to zonal flows and zonal density, respectively.

\begin{figure}
\begin{center}
\begin{tabular}{ll}
\includegraphics[width=0.5\linewidth]{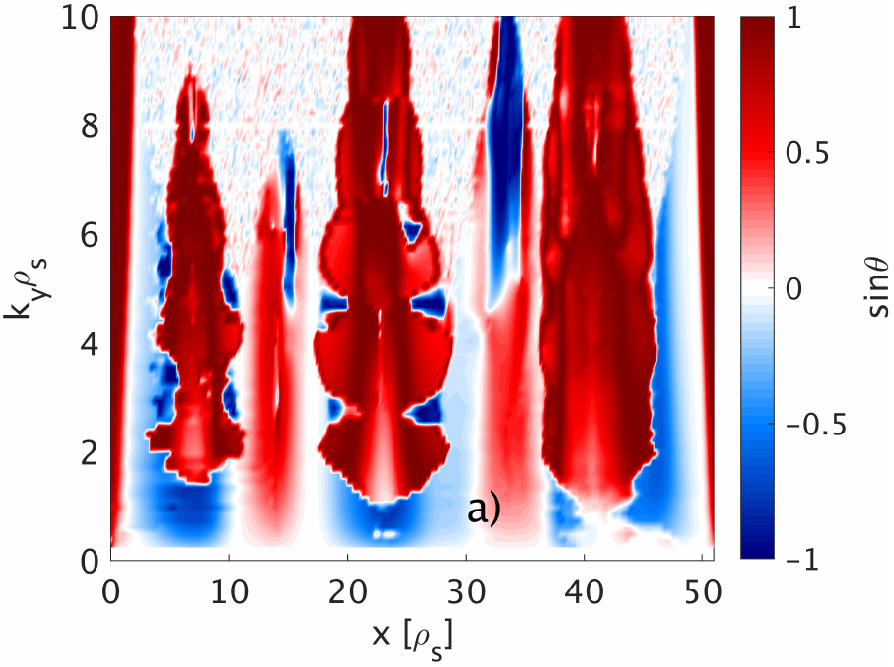} & \includegraphics[width=0.5\linewidth]{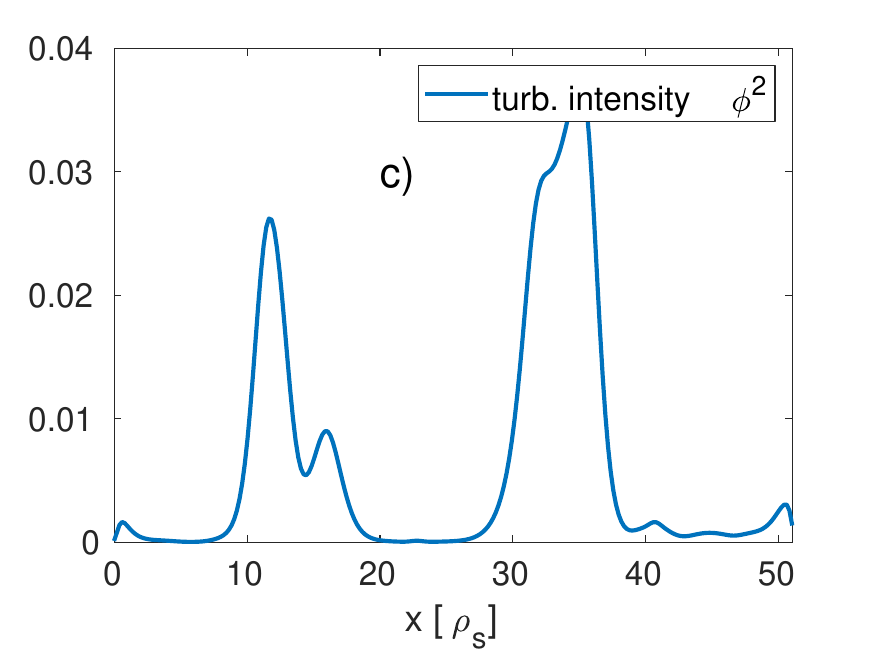} \\\includegraphics[width=0.5\linewidth]{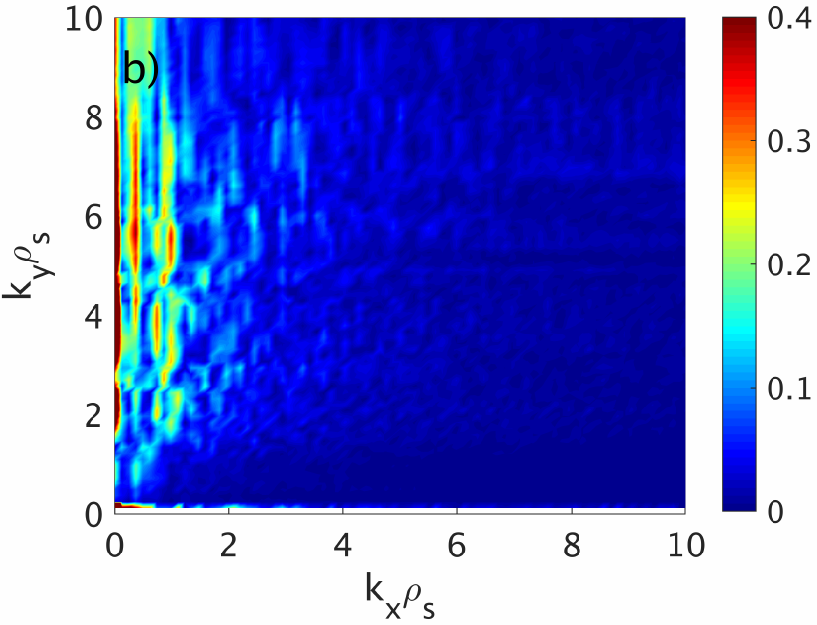}
\end{tabular}
\caption{Evidence of radial modulation of the transport crossphase: a) time-averaged nonlinear crossphase spectrum $\sin\theta_k$, v.s. poloidal wavenumber $\ky$ and radial position, b) radial Fourier transform of the crossphase, and c) radial profile of the turbulence intensity $\langle \tilde \phi^2 \rangle$. The adiabaticity parameter is $\anorm=2$.}
\label{fig-radmod}
\end{center}
\end{figure}

\begin{figure}
\begin{center}
\includegraphics[width=0.5\linewidth]{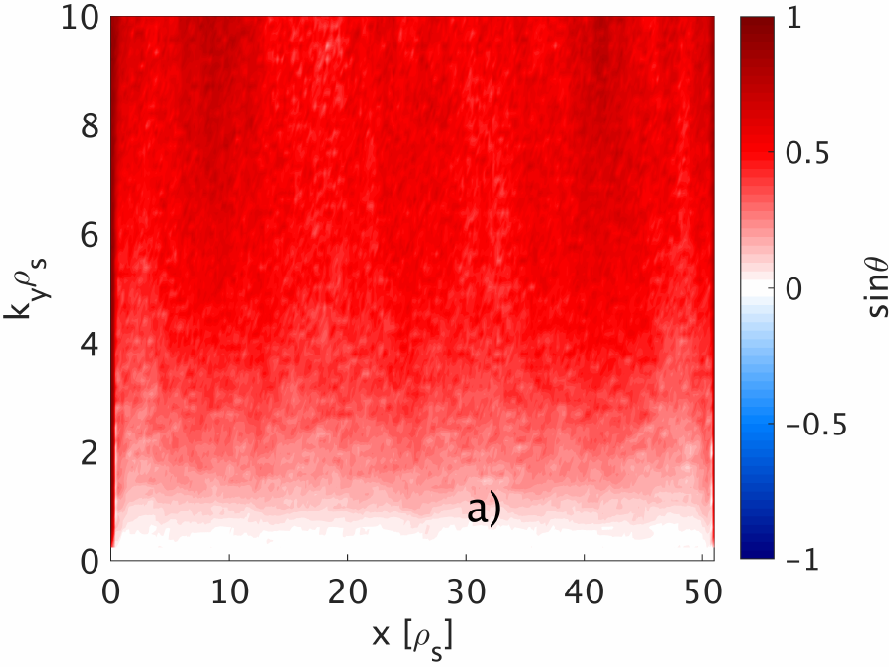}\quad\includegraphics[width=0.4\linewidth]{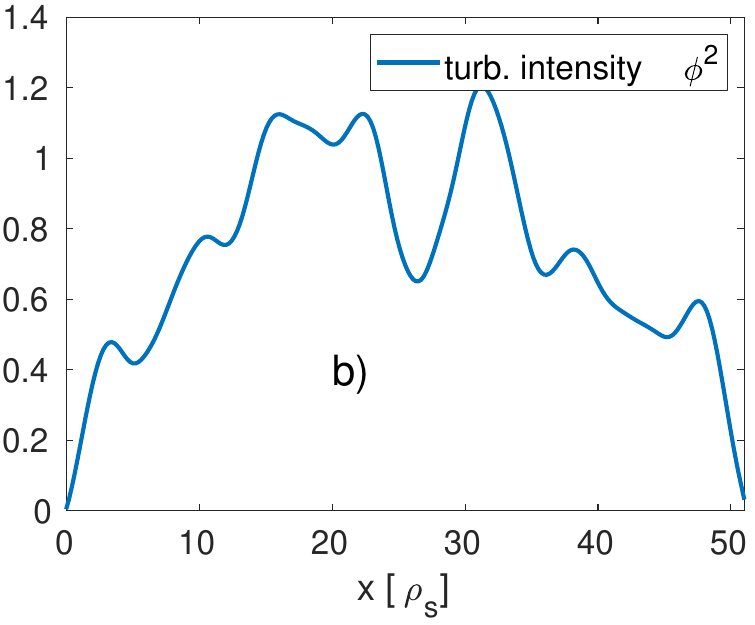}
\caption{case with artificially-suppressed zonal flows: a) time-averaged nonlinear crossphase spectrum, v.s. wavenumber $\ky$ and radial position, and b) radial profile of the turbulence intensity $\langle \tilde \phi^2 \rangle$. Other parameters are the same as in Fig.\ref{fig-radmod}.}
\label{fig-radmod-nozf}
\end{center}
\end{figure}

\begin{figure}
\begin{center}
\includegraphics[width=0.5\linewidth]{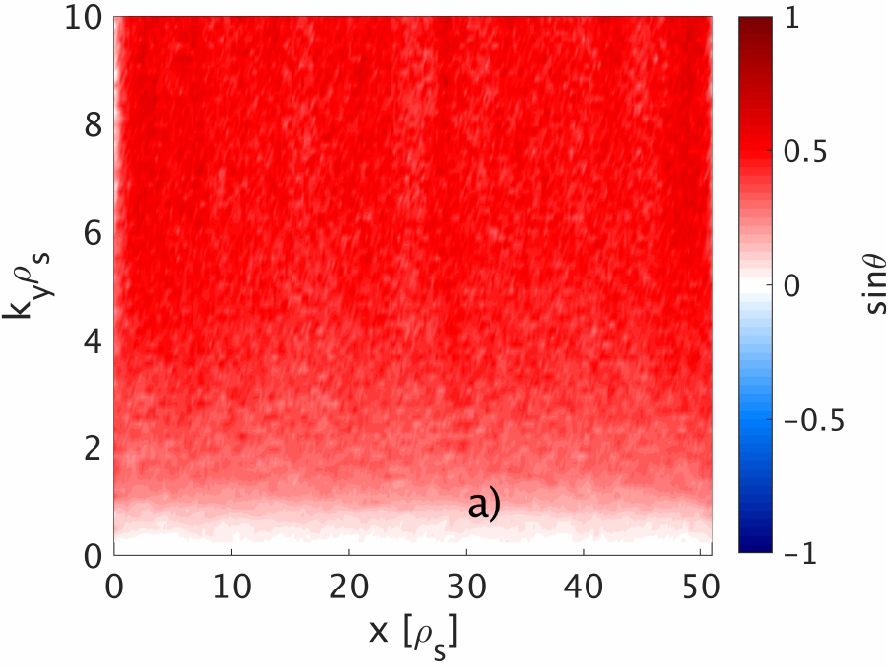}\includegraphics[width=0.4\linewidth]{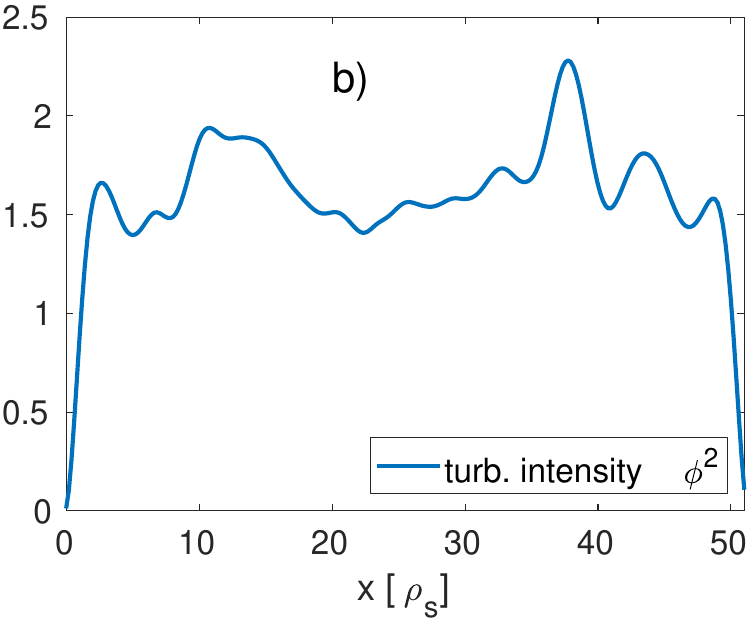}
\caption{case with artificially suppressed zonal flows and zonal density: a) time-averaged nonlinear crossphase spectrum, v.s. wavenumber $\ky$ and radial position, and b) Radial profile of the turbulence intensity $\langle \tilde \phi^2 \rangle$. Other parameters are the same as in Fig.\ref{fig-radmod}.}
\label{fig-radmod-nozfnozn}
\end{center}
\end{figure}

In the saturated state, the nonlinear transport crossphase $\cp = \arg ( n_k^* \phi_k)$, averaged over time, is shown v.s. poloidal wavenumber $\ky$ and radial direction $x$ [Fig.\ref{fig-radmod}a]. Here, $\arg(z)$ denotes the argument of the complex $z$. A modulation pattern is clearly observed in the radial direction $x$ [Fig.\ref{fig-radmod}a]. We stress that this is the first time that such a radial modulation of transport crossphase has ever been observed in numerical simulations. The reason is that gyrokinetic simulations tend to focus on the poloidal wavenumber $\ky$ dependence \cite{Told2013}. This nonlinear modulation of the crossphase can act as a stabilization of turbulence, even when zonal flows are artificially suppressed [Fig.\ref{fig-radmod-nozf}a]. For comparison, the reference case with suppressed zonal flows and zonal density is also shown  [Fig.\ref{fig-radmod-nozfnozn}a]. The associated level of turbulence, i.e. the turbulence intensity profile $\langle \tilde\phi^2 \rangle$ is shown for the case with zonal flows and zonal density [Fig.\ref{fig-radmod}b], for the case with artificially-suppressed zonal flows [Fig.\ref{fig-radmod-nozf}b], and for the reference case of zonal flows and zonal density both artificially-suppressed [Fig.\ref{fig-radmod-nozfnozn}b]. One observes that the turbulence level is the highest in the latter case $\langle \tilde \phi^2 \rangle \sim 1.5$, as expected. In the case with both zonal flows and zonal density [Fig.\ref{fig-radmod}b], the turbulence level is strongly suppressed $\langle \tilde \phi^2 \rangle \sim 2 \times 10^{-2}$, consistent with the standard ZF-induced eddy-shearing paradigm. However, even with artificially-suppressed zonal flows [Fig.\ref{fig-radmod-nozf}b], the average turbulence level $\langle \tilde \phi^2 \rangle \sim 0.8$ is lower than for the reference case with artificially-suppressed zonal flows and zonal density [Fig.\ref{fig-radmod-nozfnozn}b]. This shows that the turbulence is partly-suppressed by zonal density corrugations, qualitatively consistent with the extended wave-kinetic model (\ref{wke1},\ref{zv1},\ref{zn1}).

\begin{figure}
\begin{center}
\includegraphics[width=0.4\linewidth]{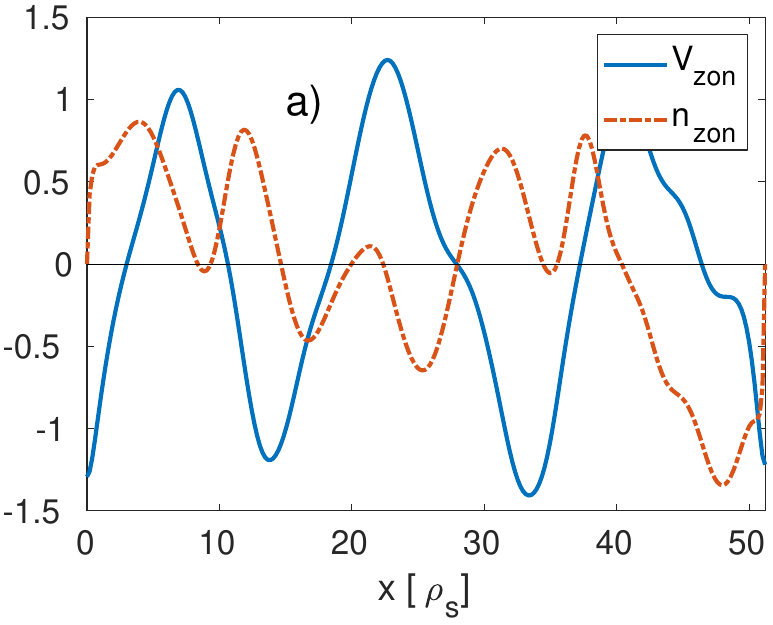}\includegraphics[width=0.4\linewidth]{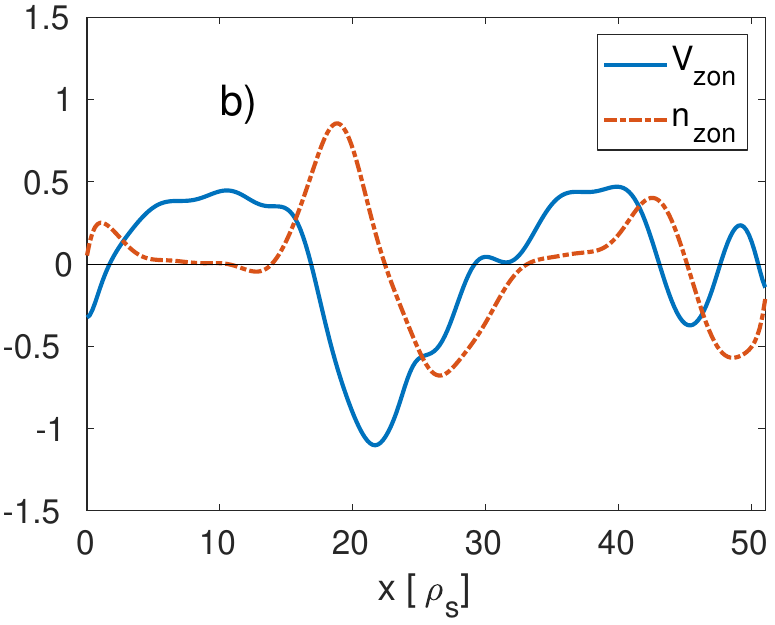}
\caption{Time-averaged profiles of zonal flows (blue) and zonal density (red), for a) $\anorm=2$ and b) $\anorm=10$. Other parameters are the same as in Fig.\ref{fig-radmod}.}
\label{fig-prof}
\end{center}
\end{figure}

\begin{figure}
\begin{tabular}{cc}
\includegraphics[width=0.5\linewidth]{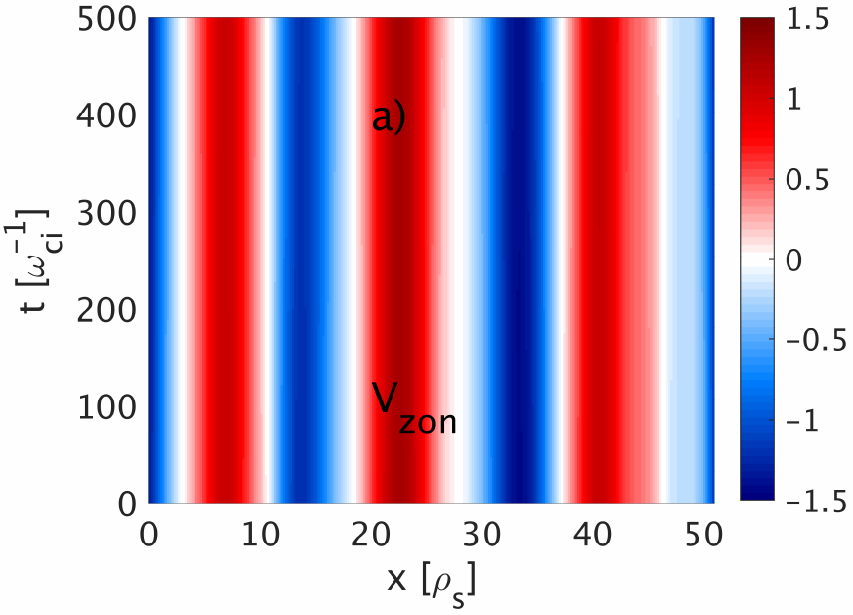} & \includegraphics[width=0.5\linewidth]{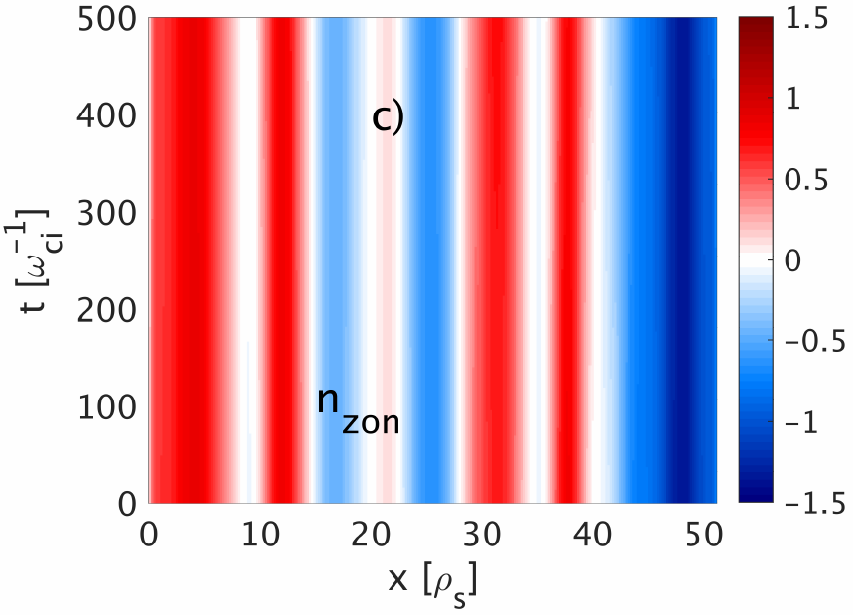} \\
\includegraphics[width=0.42\linewidth]{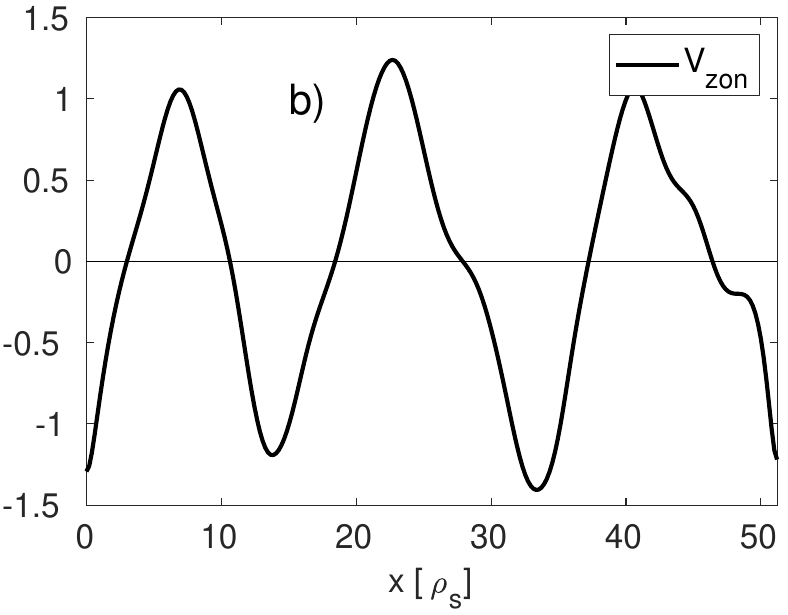} & \includegraphics[width=0.42\linewidth]{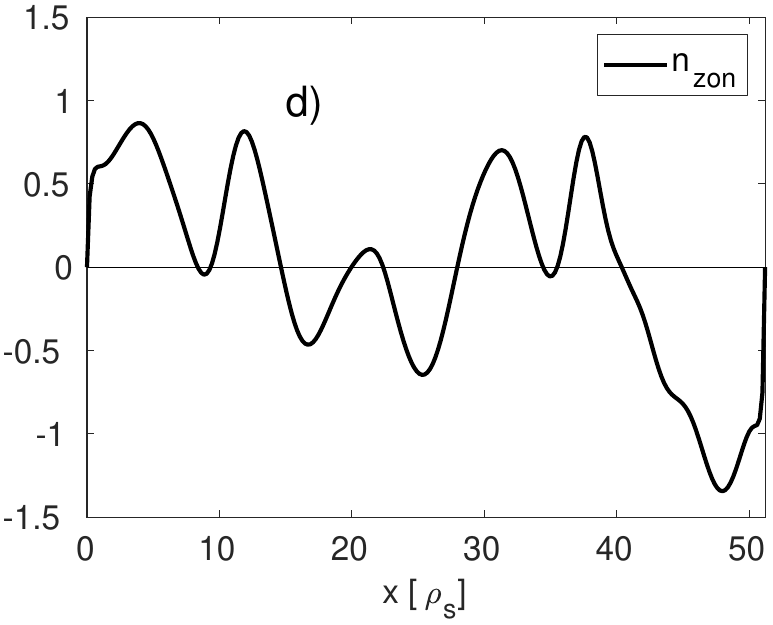}
\end{tabular}
\caption{a) Zonal flow spatio-temporal dynamics and b) time-averaged zonal flow profile, and c) zonal density spatio-temporal dynamics, d) time-averaged zonal density profile. The adiabaticity parameter is $\anorm=2$.}
\label{fig-xt}
\end{figure}



\begin{figure}
\begin{center}
\includegraphics[width=0.5\linewidth]{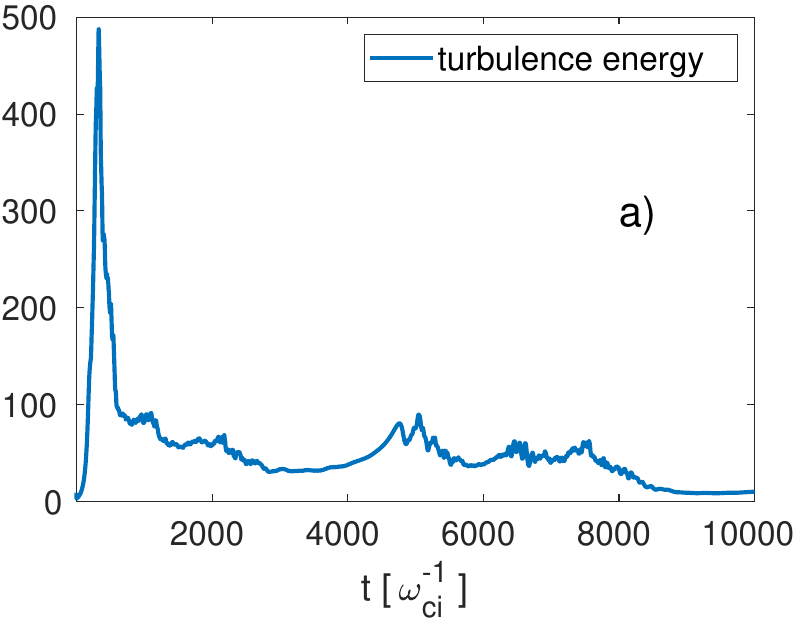}\includegraphics[width=0.5\linewidth]{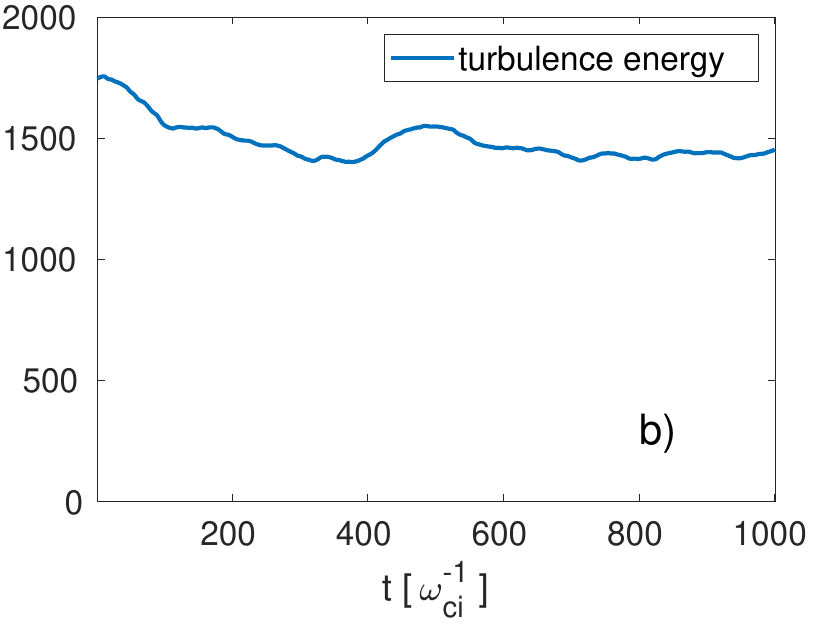}
\caption{Time-series of the turbulence energy: a) with zonal flows present and b) with artificially-suppressed zonal flows. The parameters are the same as in Fig.\ref{fig-radmod}. Case b) is shown after the turbulence has reached the saturation regime.}
\label{fig-ene}
\end{center}
\end{figure}

\begin{figure}
\begin{tabular}{ll}
\includegraphics[width=0.5\linewidth]{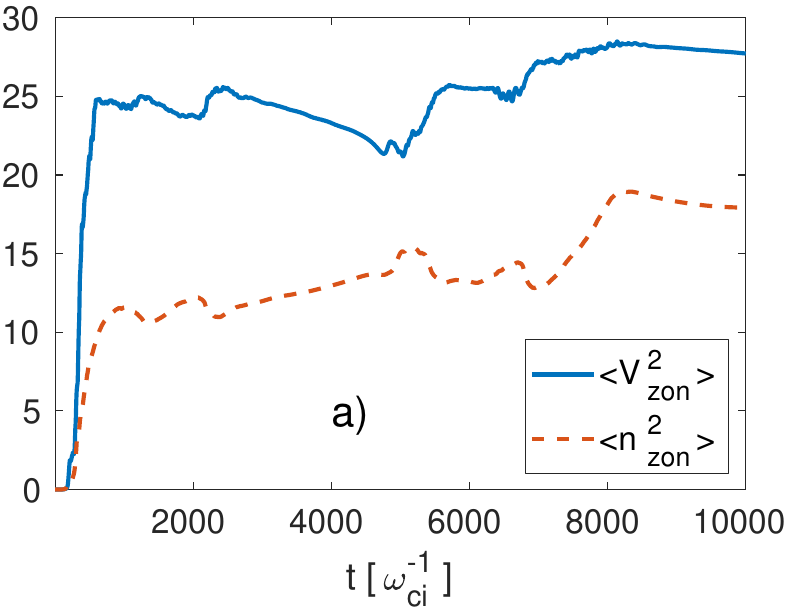} & \includegraphics[width=0.5\linewidth]{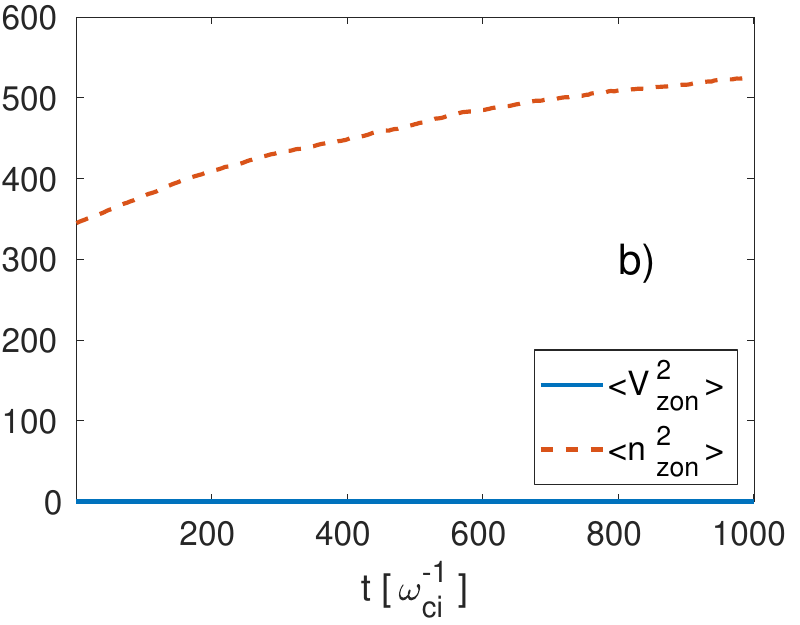} \\
\includegraphics[width=0.5\linewidth]{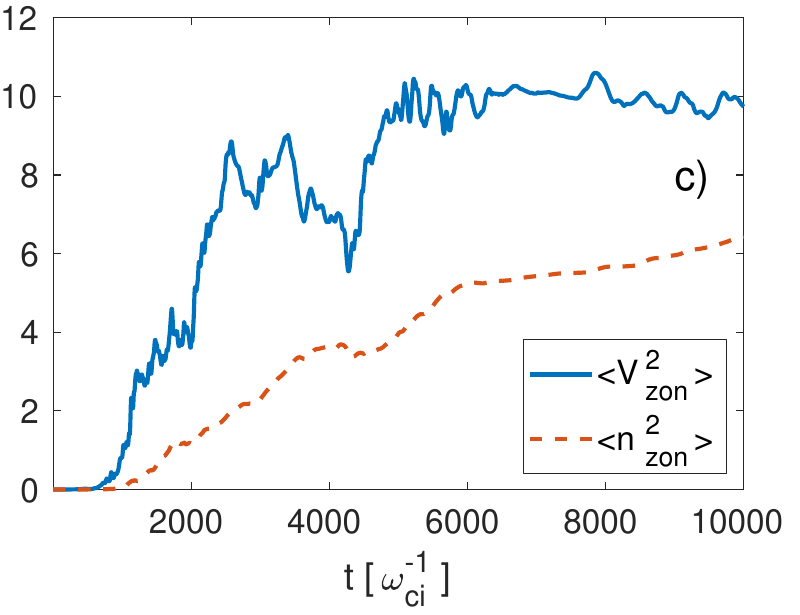}
\end{tabular}
\caption{Time-series of the energy of zonal flows (full-line) and zonal density (dash): a) with zonal flows present and $\anorm=2$, b) with artificially-suppressed zonal flows and $\anorm=2$, and c) with zonal flows present and $\anorm=10$. The parameters are the same as in Fig.\ref{fig-radmod}, except c) for which $\anorm=10$.}
\label{fig-enezf}
\end{figure}

The profiles of zonal flows and zonal density, averaged over time,  are shown [Fig.\ref{fig-prof}a,b] for two values of the adiabaticity parameter $\anorm=2$ and $\anorm=10$. In the low-adiabaticity regime $\anorm=2$, it is apparent from Fig.\ref{fig-prof}a that the zonal density profile evolves on a much-smaller scale than that of zonal flows. This is somewhat surprising and not intuitive, since in the litterature, zonal flows and zonal density are described as two components of `zonal modes' with the same wavenumber ${\bf q}=q_x \hat x$. Hence, if zonal modes were really describable in this way, we would naively expect that they would have approximately the same scale, i.e the same dominant wavenumber. The fact that they have widely different scale - and hence different dominant wave number - points to the intrinsically nonlinear nature of these fields. The different scale between zonal flows and zonal density in drift-wave turbulence was first observed in Ref. \cite{KimAn2019}. This difference of scale between zonal flows and zonal density can be interpreted - in the low adiabaticity regime - as a \emph{decoupling} between different transport channels, i.e. vorticity transport v.s. particle transport. For a more sophisticated model, this may have implications for the important phenomenon of \emph{transport decoupling} between particle transport v.s. heat transport as observed e.g. in the improved confinement mode (I-mode) with high energy confinement but low particle confinement \cite{Hubbard2016}, where the crossphase modulations may play a crucial role, but more work needs to be done to confirm this picture. At high adiabaticity $\anorm=10$, however, i.e. for nearly Boltzmann electrons, the radial scale of zonal density becomes comparable to that of zonal flows [Fig. \ref{fig-prof}b].
The spatio-temporal dynamics of zonal flows and zonal density is shown, for the case $\anorm=2$ [Fig.\ref{fig-xt}]. The turbulence energy evolution [Fig.\ref{fig-ene} a,b], and the energy of zonal flows and zonal density [Fig.\ref{fig-enezf} a-c] are also shown.


\subsection{Comparison with the extended wave-kinetic model \& with JFT-2M experimental data}

\subsubsection{Comparison with the extended wave-kinetic model}
\begin{figure}
\begin{center}
\includegraphics[width=0.5\linewidth]{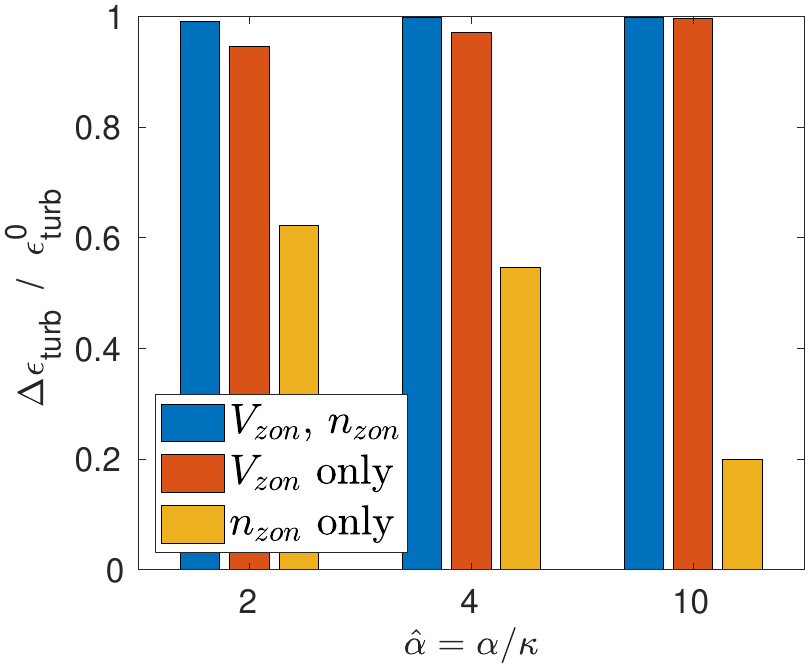}
\caption{Zonal efficiency $\Upsilon = \Delta \epsilon_{\rm turb} / \epsilon_{\rm turb}^0$ for several values of adiabaticity parameter $\anorm=2$, $\anorm=4$ and $\anorm=10$, and for different cases: blue: with zonal flows and zonal density, red: with artificially-suppressed zonal density, and yellow: with artificially-suppressed zonal flows.}
\label{fig-effzon-alpha}
\end{center}
\end{figure}

In this section, qualitative comparison will be made between aspects of the extended wave-kinetic model (\ref{wke1},\ref{zv1},\ref{zn1}) and associated reduced 1D model (\ref{wke2},\ref{zv2},\ref{zn2}) on one hand, and the numerical simulations of the full model on the other hand.
First, the extended wave-kinetic model predicts that zonal flows and zonal density corrugations both play a role in  transport suppression. Hence, one should not be dominant over the other. This is consistent with the time-average profile of Fig.\ref{fig-prof} which shows them to be roughly of the same magnitude. The reduced model also predicts the radial modulation $\Delta \cp(x,t)$ of the transport crossphase, which is confirmed in Fig.\ref{fig-radmod}a. When zonal flows are artificially suppressed, the amplitude of the crossphase modulation seems to decrease [Fig.\ref{fig-radmod}a], and crossphase modulations vanish completely when both zonal flows and zonal density are artificially suppressed Fig.[\ref{fig-radmod-nozfnozn}a]. This shows that this radial modulation is a nonlinear phenomenon and is not predicted by quasi-linear theory which uses the linear `$i \delta$' prescription for the electron response. One may ask: What is the qualitative effect of zonal density v.s. zonal flows in suppressing the turbulence?
To investigate this, we compare the turbulence level, i.e. the time-averaged turbulence energy at saturation, for different cases. It is convenient to introduce a normalized indicator: the \emph{zonal efficiency} $\Upsilon$ (\%), defined as:
\begin{equation}
\Upsilon = \frac{\Delta \epsilon_{turb}}{\epsilon_{turb}^0},
\label{def-effzon}
\end{equation}
where $\Delta \epsilon_{turb} = | \epsilon_{turb} - \epsilon_{turb}^0 |$, $\epsilon_{turb}$ denotes the time-averaged turbulence energy at saturation, and $\epsilon_{turb}^0$ is its reference value when both zonal flows and zonal density are artificially-suppressed. \\
The zonal efficiency (\ref{def-effzon}) indicates how strongly different zonal structures are able to suppress turbulence. It is shown for different cases [Fig.\ref{fig-effzon-alpha}]. For $\anorm=2$, one observes that the case with zonal flows and zonal density has a zonal efficiency of $\Upsilon \sim 99.1\%$ close to 100\%, corresponding to almost totally suppressed turbulence. The case with artificially-suppressed zonal density has a zonal efficiency of $ \Upsilon \sim 94.4\%$, thus lower than the case with both zonal flows and zonal density present, although not by a large margin. However, the most interesting case is the one with zonal density alone, i.e. with artificially-suppressed zonal flows, with a zonal efficiency of $\Upsilon \sim 61.2\%$,which is still large. This shows that zonal density corrugations may play an important role in turbulence suppression in some regimes. For $\hat \alpha=4$, the zonal efficiency is qualitatively similar to the case $\hat \alpha=2$, although there are small quantitative differences, as zonal density effects become weaker. This trend continues for $\anorm=10$, where zonal density effects become negligeable: with zonal density alone, in this case, zonal efficiency is $\Upsilon \sim 20\%$ only.


\subsubsection{Qualitative comparison with JFT-2M experimental data}
We also compare the theory with experimental data from previous JFT-2M experimental observations of limit-cycle oscillations (LCO) during the L-H transition \cite{Kobayashi2014}. Data from the heavy ion beam probe (HIBP) shows qualitative features resembling the zonal density corrugations of our model. It should be noted that these density corrugations are interpreted to be induced by the turbulence spreading, where the turbulence clump and density gradient perturbation simultaneously travel \cite{Kobayashi2014}. We leave detailed comparison for future work.
Data from Ref. \cite{Kobayashi2014} shows a slow spatial modulation of the HIBP profile, a proxy for electron density.
The sound Larmor radius is $\rho_s \sim \rho_i \sim 1.2 mm$ in this experiment, where $T_e \sim T_i$ is assumed and $\rho_i$ is the ion Larmor radius. Ref. \cite{Kobayashi2014} estimated the radial wavenumber of the LCO as $q_r \sim 25 m^{-1}$ and that of the microturbulence as $k_r \sim 10^2 m^{-1}$. This gives $q_r \rho_s \sim 0.03$ and $k_r \rho_s \sim 0.12$, hence $q_r \rho_s \ll k_r \rho_s$, consistent with a slow radial modulation.


\section{Discussion}
The wave-kinetic model Eqs. (\ref{wke1},\ref{zv1},\ref{zn1}) is an extension of the well-know wave-kinetic equation \cite{MattorDiamond1994, Bretherton1969}, to self-consistently include the physics of the transport crossphase. We can compare this model with that of Ref. \cite{Sasaki2018}. The main difference is the presence of the nonlinear part of the growth-rate in our model, second term on the r.h.s. of Eq. (\ref{wke1}), which can be traced to the convective $E \times B$ nonlinearity. This couples to the dynamics of zonal density corrugations, providing a new feedback loop which is absent in the standard wave-kinetic equation. In Ref. \cite{Sasaki2018}, the wave-kinetic model is solved numerically in the extended phase-space ($x, k_x$). It shows a complex interplay between turbulence and zonal flows that lead to nonlinear structures (patterns), associated to the `trapping' of turbulence wave-packets in the troughs of zonal flows. This is beyond the scope of this article and left for future work. Instead, we provides evidence of the validity of the model by using fluid simulations in real space.
Let us now discuss the zonal density generation mechanism, Eqs. (\ref{zn1}) and (\ref{zn2}). Ref. \cite{LangParkerChen2008} showed that CTEM turbulence can saturate via nonlinear generation of zonal density. We may compare the zonal density drive mechanism described by equation (8) in \cite{LangParkerChen2008}, as this should be model-independent. Our Eq. (\ref{zn1}) differs from the one in \cite{LangParkerChen2008}, since we show that energy is conserved between turbulence and zonal density, whereas the analysis in \cite{LangParkerChen2008} is valid only for the initial exponential growth of the modulational instability. The fluid model (\ref{hw1},\ref{hw2}) could possibly be extended to CTEM, where the drive of zona density structures seems to play a crucial role \cite{QiMJChoiLeconte2022,LeconteMassonQi2022}.

There are limitations to our model. First, the model assumes cold ions, $T_i \ll T_e$, and hence does not contain finite-ion Larmor radius (FLR) effects, although it includes ion inertia ($\rho_s$) effects. It is thus not directly applicable to the important ion-temperature-gradient driven mode (ITG). Second, electron temperature gradient effects ($\eta_e$) are neglected. This is beyond the scope of this article and left for future work, where we plan to investigate the possible decoupling between particle transport and thermal transport (transport decoupling).

\section{Conclusion}
In this work, we derived the extended wave-kinetic equation (\ref{wke1}), self-consistently coupled to the dynamics of zonal flows and zonal density corrugations. The latter may be a missing piece in the understanding of turbulence in fusion plasmas. The theory can be summarized as follows: Turbulent fluctuations self-organize to generate quasi-stationary radial modulations $\Delta \cp(r,t)$ of the transport crossphase $\cp$ between density fluctuations and potential fluctuations. This results in turbulent particle flux modulations $\tilde \Gamma(r,t)$. The radial modulation of particle flux nonlinearly drive zonal  corrugations of the density profile via a modulational instability. In turn, zonal density corrugations regulate the turbulence via nonlinear damping of the fluctuations.
The main findings of this work are:
i) The present theory takes into account the convective $E \times B$ nonlinearity, and thus goes beyond the well-known `$i \delta$' quasi-linear approximation,
ii) This nonlinear mechanism conserves energy between turbulence and zonal density. Since zonal density is a radial mode ($m=0,n=0$), with $m$ and $n$ the poloidal  and toroidal mode numbers, it cannot drive transport and thus provides a benign reservoir of energy for the turbulence, and
iii) In fluid simulations of collisional drift-wave turbulence, the radial modulation of the transport crossphase and associated staircase profile structure have been confirmed to partly stabilize the turbulence.

\section*{Acknowledgments}
M.L. would like to thank J.M. Kwon, Lei Qi, I. Dodin, Hongxuan Zhu, T. Stoltzfulz-Dueck and M.J. Pueschel for helpful discussions. M.L was supported by R\&D Program through Korean Institute for Fusion Energy (KFE) funded by the Ministry of Science and ICT of the Republic of Korea (No. KFE-EN2341-9).


\section*{Appendix: derivation of the reduced model}
Here, we detail the derivation of the reduced model (\ref{wke1},\ref{zv1},\ref{zn1}). 
Linearizing Eq. (\ref{hw1}) yields:
\begin{equation}
n_k^L = \left[ 1 - i  \frac{\dia - \wk^L}{\alpha} \right] \phi_k
\end{equation}
which provides the linear density response:
\begin{equation}
n_k^L = ( 1 - i \cp^0 ) \phi_k,
\end{equation}
with $\cp^0= (\dia - \wk^L)/ \alpha$ the linear transport crossphase.
Moreover, the first-order modulation of Eq. (\ref{hw1}) yields:
\begin{equation}
\Delta n_k = i (\dia - \wk^L) \frac{ \ky \nabla_x \zn }{ \dia \alpha} \phi_k
\end{equation}
which provides the nonlinear correction to the density response:
\begin{equation}
\Delta n_k = - i \Delta \cp \phi_k,
\end{equation}
with $\Delta \cp = - \ky \nabla_x \zn/ \alpha$ the crossphase radial modulation.
Hence, in the weak-turbulence approximation, the nonlinear density response is:
\begin{eqnarray}
n_k & \simeq & n_k^L + \Delta n_k \\
 & \simeq & [1- i (\cp^0 + \Delta \cp) ] \phi_k
\end{eqnarray}

Substracting Eq. (\ref{hw2}) from Eq. (\ref{hw1}) yields the conservation of potential vorticity i.e. gyrocenter ion density:
\begin{equation}
\frac{\dif}{\dif t} (n -\nabla_\perp^2 \phi) + v_{*0} \frac{\dif \phi}{\dif y} + {\bf v}_E . \nabla (n - \nabla_\perp^2 \phi) = 0
\end{equation}
Using the weak-turbulence approximation \cite{ZhouZhuDodin2019}, this can be written:
\begin{equation}
\frac{\dif}{\dif t} (n_k + k_\perp^2 \phi_k) + i \dia \phi_k + {\bf V}_{zon} . \nabla (n_k + k_\perp^2 \phi_k) = 0,
\end{equation}
where ${\bf V}_{zon} = \hat z \times \nabla \phi_{zon}$ denotes zonal flows.
After some algebra, this can be written in the form of a Schrodinger-like equation:
\begin{equation}
i \frac{\dif}{\dif t} (R_k + k_\perp^2) \phi_k = \Big[ \dia + \ky U (R_k + k_\perp^2 \Big] \phi_k ,
\end{equation}
with $R_k = 1- i (\cp^0 + \Delta \cp)$, and $U = V_{zon}$. \\
Assuming $|\dif_t \Delta \cp| \ll |\dif_t \phi_k|$, one obtains:
\begin{equation}
i \frac{\dif \phi_k}{\dif t}   = \Big[ \frac{\dia}{1 + k_\perp^2 - i (\cp^0 + \Delta \cp) } + \ky U(x,t)  \Big] \phi_k ,
\end{equation}
Using the approximation $|\cp^0|, |\Delta \cp| \ll 1$, this reduces to:
\begin{equation}
i \frac{\dif \phi_k}{\dif t}   = H_H \phi_k +i H_A \phi_k,
\end{equation}
where  $H_H = \wk + \ky \zv(x,t)$ and $H_A = \gamma_k^0 + \Delta \gamma_k(x,t)$ denote the Hermitian and anti-Hermitian parts of the `Hamiltonian', respectively.
Here, $\gamma_k^0 = \dia \cp^0$ the linear growth-rate associated to the linear crossphase $\cp^0$, and $\Delta \gamma_k = \dia \Delta \cp(x,t)$ the nonlinear growth-rate associated to the nonlinear part $\Delta \cp(x,t) \sim \ky N(x,t)$ of the crossphase. Here $N(x,t)$ denotes the zonal density gradient.

Following Ref.\cite{ZhouZhuDodin2019}, one obtains the following wave-kinetic equation:
\begin{equation}
\frac{\dif W_k}{\dif t} + \{ H_H, W_k \}  = 2 H_A W_k,
\end{equation}
where here $\{ \cdot, \cdot \}$ denotes the Poisson bracket in $(k_x, x)$ extended phase space.

\end{document}